\documentclass{aa}
\usepackage{aabib}
\usepackage{epsf}
\begin{document}
\thesaurus{07(08.01.3;08.03.1;09.04.1;13.09.4)}
\title{Infrared properties of SiC particles} 
\author{H. Mutschke \inst{1}
\and A.C. Andersen \inst{2}
\and D. Cl\'ement \inst{1} 
\and Th. Henning   \inst{1} 
\and G. Peiter \inst{3}}

\institute{Astrophysikalisches Institut und Universit\"{a}ts-Sternwarte (AIU), 
Schillerg\"a\ss{}chen 3, D-07745 Jena, Germany
\and
Astronomisk Observatorium, Juliane Maries Vej 30, 
          DK--2100 Copenhagen, Denmark 
\and
Institut f\"ur Physikalische Chemie, Lessingstra\ss{}e 10, 
D-07743 Jena, Germany}
\offprints{mutschke@astro.uni-jena.de}
\date{Received date; accepted date}
\maketitle
\markboth{Infrared properties of SiC}{Mutschke et al.}

\begin{abstract}
We present basic laboratory infrared data on a large number of SiC 
particulate samples, which should be of great value for the 
interpretation of the 11.3~$\mu$m feature observed in the spectra 
of carbon-rich stars. The laboratory spectra show a wide variety 
of the SiC phonon features in the 10-13~$\mu$m wavelength range, 
both in peak wavelength and band shape. The main parameters determining 
the band profile are morphological factors as grain size and shape 
and, in many cases, impurities in the material. We discovered the 
interesting fact that free charge carriers, generated e.g. by nitrogen 
doping, are a very common characteristics of many SiC particle samples. 
These free charge carriers produce very strong plasmon absorption in the 
near and middle infrared, which may also heavily influence the 
10-13~$\mu$m feature profile via plasmon-phonon coupling. 

We also found that there is no systematic dependence of the band profile 
on the crystal type ($\alpha$- vs. $\beta$-SiC). This is proven both 
experimentally and by theoretical calculations based on a study of the 
SiC phonon frequencies. Further, we give optical constants 
of amorphous SiC. We discuss the implications of the new laboratory 
results for the interpretation of the spectra of carbon stars.


\end{abstract}

\section{Introduction}

Thermodynamical equilibrium calculations performed by Friedemann (1969a,b) 
\nocite{Friedemann69a,Friedemann69b} and \cite*{Gilman69} suggested that 
silicon carbide (SiC) particles can form in the mass outflows from carbon 
stars. First empirical evidence for the presence of such particles came from 
infrared spectroscopy performed by \cite*{Hackwell72} and \cite*{Treffers74}. 
The latter authors interpreted the detected relatively broad 11.3\,$\mu$m 
feature as the emission of small SiC particles between the transverse and 
longitudinal optical phonon frequencies. More empirical material was 
accumulated by the IRAS mission (\cite{Little-Marenin86,Goebel95}) and
recent ground-based observations with UKIRT (\cite{Speck97}). The UKIRT
data showed that the feature cannot only appear in emission, but that there
are a few sources with an absorption feature. 

An ultimate proof for the formation of SiC grains in carbon-rich stellar 
atmospheres was the detection of isotopically anomalous SiC grains in 
primitive meteorites (\cite{Bernatowicz87}). The s-process isotopic signature 
identified in the majority of the meteoritic SiC grains can be produced in 
the atmospheres of thermally pulsing low-mass AGB stars (\cite{Gallino94}). 
Most of these presolar SiC grains seem to be large with equivalent 
spherical diameters between 0.3 and 3\,$\mu$m (\cite{Amari94}), but also very 
large grains up to 26\,$\mu$m are present. Such large grains can probably 
only be formed during extreme mass loss phases or in clumps of high density 
(\cite{Bernatowicz96}). So far there is no spectroscopic evidence for SiC 
grains in the diffuse interstellar medium. This fact is not yet satisfactorily 
explained although core-mantle, inhomogeneous, or very large grains may
offer a solution (\cite{Ossenkopf92,Kozasa96,Andersen98}). 

The 11.3~$\mu$m feature in carbon star spectra displays some interesting 
variability which may offer the possibility to characterize more closely 
the composition and structure of cosmic SiC grains. \cite*{Little-Marenin86} 
analyzed 176 spectra of bright infrared sources in the IRAS LRS database 
and found for 4\% of the carbon stars a feature shifted to 
$\approx$ 11.6 $\mu$m. The majority of the features peaked at 
11.15 $\pm$ 0.10 $\mu$m with a FWHM of 1.6 $\pm$ 0.15 $\mu$m.
The statistical analysis of spectra of class 4n revealed some 
more variety with only part of the spectra peaking
at 11.3\,$\mu$m and a second peak at 11.7\,$\mu$m appearing 
when the ratio of feature to continuum decreases (\cite{Baron87}). 
A very thorough study of 718 LRS spectra of carbon stars based on 
autoclassification methods was performed by \cite*{Goebel95} who 
also found a weakening, broadening and shift (to 11.9\,$\mu$m) 
for classes of objects with a low continuum temperature. 
The ground-based data obtained by \cite*{Speck97} also show in about one 
third of the features a peak at wavelengths larger than 11.4\,$\mu$m, 
but without a strong correlation to the blackbody temperatures. 

A definite explanation of these spectral differences in terms of an 
identification of a well-defined silicon carbide material as carrier 
of the 11+~$\mu$m feature would be very valuable, because it 
could provide important information concerning grain condensation 
mechanisms and the conditions within circumstellar grain-forming regions. 
Unfortunately, this task is complicated by (i) the presence of only one 
relatively broad band, (ii) some uncertainty in the location of the 
continuum due to nearby molecular bands (\cite{Hron98}) (iii) possible 
contribution from carbonaceous material to the emission, (iv) the extreme 
sensitivity of the SiC band profile to particle size and shape effects, 
(v) the appearance of many different SiC crystal types (polytypes), 
(vi) the possibility of disordered structures, and (vii) the influence 
of material impurities on the band profile. 

Three different explanations of the observed band variety have been given 
so far. 
\cite*{Baron87} and \cite*{Goebel95} found a correlation of the apparent 
band shift to the appearance of a feature at about 8.5~$\mu$m as well as 
to an increase in the FIR flux and attributed these combined spectral changes 
to an increasing contribution by hydrogenated amorphous carbon (a:C--H). 
\cite*{Speck97} explained the different band profiles by different 
crystal types of the SiC grains and tried to fit the UKIRT spectra 
to laboratory spectra (see below). \cite*{Papoular98} showed that 
also different grain shapes can produce the observed variation in 
band shape. 

Considering this complexity of the identification problem (see also 
\cite{Sandford96}), detailed laboratory investigations on the 
infrared spectrum of SiC particles unfortunately are scarce 
both in the physical/chemical and the astronomical literature. 
Most popular among astronomers are the studies by 
\cite*{Stephens80} on laser-produced $\beta$-SiC condensates, 
\cite*{Friedemann81} on commercially available $\alpha$-SiC, 
and \cite*{Borghesi85} on commercially produced
$\alpha$- and $\beta$-SiC. More recently, \cite*{Papoular98}
investigated two samples of $\beta$-SiC powders, one produced by laser
pyrolysis and one which was commercially available. In the earlier
papers, influential factors beyond the polytype remained mostly
unexplored. Especially, the decisive influence of grain shape and 
size as well as of the surrounding medium (matrix) on the band profile, 
which was already discussed by \cite*{Treffers74} and later investigated 
in detail by \cite*{Bohren83}, has not been treated correctly or even 
ignored by the majority of the authors. \cite*{Papoular98} urgently brought 
this topic back to mind and presented an exact method for the correction 
of matrix effects. 

As already pointed out, the available experimental studies with 
a total of about 15 samples are by no means systematic. Moreover, 
the lack of good laboratory results led to some confusion 
concerning the influence of the crystal type on the band profile. 
The goal of our paper is to present experimental data on $\alpha$-SiC,
$\beta$-SiC, and amorphous SiC and to investigate experimentally size, 
shape, and matrix effects (Sect.\,3 and 4). To give the discussion on 
the optical properties of different SiC polytypes a more exact foundation, 
we summarize in Sect.\,2 basic facts on SiC bulk crystals and thin films 
and calculate theoretical spectra of small SiC grains for different polytypes. 
In Sect.\,5 we discuss in which way the new data can serve the observers for 
an evaluation of their data in terms of the SiC hypothesis.

\section{Basic facts on structural and infrared optical
properties of SiC}
\subsection{Crystal structure of SiC}
\begin{figure}
\centering
\leavevmode
\epsfxsize=0.9
\columnwidth 
\epsfbox{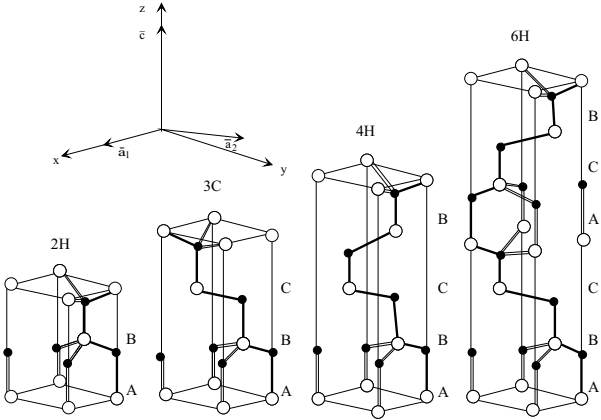}
\caption[]{Primitive hexagonal unit cells of the most simple SiC polytypes. 
Si atoms are represented by open circles, C atoms by filled circles. Bilayers of 
the three possible positions in projection with the c-axis are labeled 
by the letters A, B, and C. The Si-C bonds in the (11\={2}0) plane indicating 
the relative shifts of the bilayers are represented by heavy solid lines. 
The figure has been reproduced from \cite*{Kaeckell94}. }
\label{poly}
\end{figure}

The crystal structure of SiC shows pronounced polytypism which means 
that there exist a number of possible crystal types differing 
in only one spatial direction (see Fig.\,\ref{poly}). The basic units 
from which all polytypes are 
built are Si-C bilayers with a three-fold symmetry axis, in which the 
Si- and C-atoms are closely packed. These bilayers are stacked on top of 
each other while they are laterally shifted by 1/$\sqrt{3}$ of the Si--Si 
or C--C atomic distance in the layer either in the [\={1}100]- or in the 
opposite direction. Hence, each Si atom is tetrahedrally 
surrounded by four C atoms and vice versa. If all shifts occur in the 
same direction, then an identical position of the bilayer in the
projection along the hexagonal axis is reached after three stacking steps. 
The resulting structure is of cubic symmetry and because of the three-step 
stacking period this polytype is called 3C (C for cubic, \cite{Ramsdell47}). 
Another name for this polytype, which is the only cubic one, is the 
often-used term $\beta$-SiC. 
The environment of each Si-C bilayer in the stack, which is determined 
by shifts in the same direction at each step, is also called cubic.

The other extreme is obtained, when the bilayers 
are shifted alternatingly in opposite directions such that, 
in projection with the hexagonal axis, every other layer has the 
same position. The lattice is then of hexagonal type, and because 
of the two-step period the polytype is called 2H. 
The environment of each Si-C bilayer which has been produced by 
the alternating shiftes is also called hexagonal. 

All other polytypes are built up by a characteristic sequence of cubic and
hexagonal Si-C bilayers, for which the 3C and 2H polytypes represent the 
limiting cases. 
All polytypes except 3C are uniaxial crystals (optical axis = c-axis) and 
belong either to the hexagonal or to the rhombohedric system. 
The most abundant polytypes besides 3C and 2H are the hexagonal 
types 6H and 4H and the rhombohedric 15R. For historical 
reasons any non-cubic polytype or mixture of polytypes is also 
designated by the term $\alpha$-SiC. 

The ratio of the numbers of hexagonal to cubic bilayers is
called hexagonality and is a very useful scaling parameter. Several properties
of the polytypes change with this parameter, e.g. structural
properties such as the lattice constants (see, e.g., \cite{Landolt82,Bechstedt97}),
the energy of the indirect band gap (\cite{Feitknecht71,Choyke90}), and
certain phonon frequencies (\cite{Nakashima87b,Hofmann94}), which will be 
discussed below. 

\subsection{Lattice vibrations of bulk SiC and thin films}

\begin{table*}[htbp] 
\centering
\caption{Hexagonality parameter h, resonance frequencies of the strong phonon modes, 
and high frequency dielectric constants of some SiC polytypes. E$_{1T}$ and E$_{1L}$ 
denote the frequencies of the TO- and LO-modes in the basal plane of the SiC lattice, 
respectively, whereas A$_{1T}$ and A$_{1L}$ denote those parallel to the principal axes.}
\begin{tabular}
[c]{ccc|c|c|c|c|c|c}\hline\hline
\textbf{Polytype} & \textbf{h(\%)} &  & \textbf{E}$_{\mathbf{1T}}$ &
\textbf{A}$_{\mathbf{1T}}$ & \textbf{E}$_{\mathbf{1L}}$ & \textbf{A}%
$_{\mathbf{1L}}$ & $\mathbf{\varepsilon}_{\infty\perp}$ & $\mathbf{\varepsilon
}_{\infty\parallel}$\\\hline
\textbf{3C} & 0 &
\begin{tabular}
[c]{c}%
(exp.,\,av.)\\
\\
(theor.)
\end{tabular}
& \multicolumn{2}{|c|}{%
\begin{tabular}
[c]{c}%
795.9$^{(3,6,7,9,11,12,13,15,18}$\\
$^{19,21,23,24,25,30,33,34,35)}$\\
795.7$^{(36)}$%
\end{tabular}
} & \multicolumn{2}{|c|}{%
\begin{tabular}
[c]{c}%
972.3$^{(3,6,7,9,11,12,13,18,19,}$\\
$^{21,23,24,25,26,30,32,33,34,35)}$\\
979.0$^{(36)}$%
\end{tabular}
} & \multicolumn{2}{|c}{%
\begin{tabular}
[c]{c}%
6.49$^{(37,40,43)}$\\
\\
6.54$^{(45)}$%
\end{tabular}
}\\\hline
\textbf{6H} & 33 &
\begin{tabular}
[c]{c}%
(exp.,\,av.)\\
\\
\\
(theor.)
\end{tabular}
&
\begin{tabular}
[c]{c}%
797.0$^{(1,}$\\
$^{11,14,16,17,20,}$\\
$^{22,29,31,34,35)}$\\
797.0$^{(36)}$%
\end{tabular}
&
\begin{tabular}
[c]{c}%
788.1$^{(1,}$\\
$^{4,5,14,17,}$\\
$^{20,22)}$\\
787.2$^{(36)}$%
\end{tabular}
&
\begin{tabular}
[c]{c}%
969.9$^{(1,}$\\
$^{4,17,20,22,}$\\
$^{23,35)}$\\
978.0$^{(36)}$%
\end{tabular}
&
\begin{tabular}
[c]{c}%
965.3$^{(1,}$\\
$^{4,5,12,17,}$\\
$^{22,29,34,35)}$\\
974.5$^{(36)}$%
\end{tabular}
&
\begin{tabular}
[c]{c}%
6.56$^{(38,39,}$\\
$^{40,41,42,44)}$\\
\\
6.58$^{(45)}$%
\end{tabular}
&
\begin{tabular}
[c]{c}%
6.72$^{(38,39,}$\\
$^{40,41,42)}$\\
\\
6.79$^{(45)}$%
\end{tabular}
\\\hline
\textbf{15R} & 40 & (exp.,\,av.) & 797.5$^{(8,34)}$ &  &  & 965.0$^{(34)}$ &
6.53$^{(38)}$ & 6.70$^{(38)}$\\\hline
\textbf{4H} & 50 &
\begin{tabular}
[c]{c}%
(exp.,\,av.)\\
\\
(theor.)
\end{tabular}
&
\begin{tabular}
[c]{c}%
796.6$^{(2,22,}$\\
$^{23,27,28,34,35)}$\\
797.6$^{(36)}$%
\end{tabular}
&
\begin{tabular}
[c]{c}%
783.6$^{(2,22)}$\\
\\
783.0$^{(36)}$%
\end{tabular}
&
\begin{tabular}
[c]{c}%
968.7$^{(22,}$\\
$^{23,35)}$\\
977.5$^{(36)}$%
\end{tabular}
&
\begin{tabular}
[c]{c}%
964.2$^{(22,}$\\
$^{28,34,35)}$\\
972.3$^{(36)}$%
\end{tabular}
&
\begin{tabular}
[c]{c}%
6.56$^{(38)}$\\
\\
6.55$^{(45)}$%
\end{tabular}
&
\begin{tabular}
[c]{c}%
6.78$^{(38)}$\\
\\
6.76$^{(45)}$%
\end{tabular}
\\\hline
\textbf{2H} & 100 &
\begin{tabular}
[c]{c}%
(exp.)\\
(theor.)
\end{tabular}
&
\begin{tabular}
[c]{c}%
799.1$^{(10)}$\\
799.5$^{(36)}$%
\end{tabular}
&
\begin{tabular}
[c]{c}%
769.8$^{(10)}$\\
770.3$^{(36)}$%
\end{tabular}
&
\begin{tabular}
[c]{c}%
974.5$^{(10)}$\\
975.9$^{(36)}$%
\end{tabular}
&
\begin{tabular}
[c]{c}%
968.4$^{(10)}$\\
965.6$^{(36)}$%
\end{tabular}
&
\begin{tabular}
[c]{c}%
6.51$^{(39)}$\\
6.51$^{(45)}$%
\end{tabular}
&
\begin{tabular}
[c]{c}%
6.84$^{(39)}$\\
6.87$^{(45)}$%
\end{tabular}
\\\hline\hline
\end{tabular}%
\vspace{0.3cm}
\newline
\scriptsize
\begin{tabular}
[c]{clclll}%
1 & \cite{Feldman68} & 16 & \cite{Melnichuk92} & 31 & \cite{Zorba96}\\
2 & \cite{Feldman68b} & 17 & \cite{Engelbrecht93} & 32 & \cite{Falkovsky97}\\
3 & \cite{Mitra69} & 18 & \cite{Hopfe93} & 33 & \cite{Gottfried97}\\
4 & \cite{Colwell72b} & 19 & \cite{Choo94} & 34 & \cite{Nakashima97}\\
5 & \cite{Klein72} & 20 & \cite{Liu94} & 35 & Peiter (unpublished)\\
6 & \cite{Olego82} & 21 & \cite{Yamanaka94} & 36 & \cite{Hofmann94}\\
7 & \cite{Olego82c} & 22 & \cite{Harima95} & 37 & \cite{Shaffer69}\\
8 & \cite{Nakashima86} & 23 & \cite{Nienhaus95} & 38 & \cite{Shaffer71}\\
9 & \cite{Mukaida87} & 24 & \cite{Sciacca95} & 39 & \cite{Powell72}\\
10 & \cite{Nakashima87b} & 25 & \cite{Zorba95} & 40 & \cite{Pikhtin77b}\\
11 & \cite{Okumura87} & 26 & \cite{Arnaud96} & 41 & \cite{Bogdanov82}\\
12 & \cite{Yugami87} & 27 & \cite{Feng96} & 42 & \cite{Ninomiya94}\\
13 & \cite{Feng88} & 28 & \cite{Hu96} & 43 & \cite{Moore95}\\
14 & \cite{Salvador91} & 29 & \cite{Perez-Rodriguez96} & 44 &
\cite{Logothetidis96}\\
15 & \cite{Yoo91} & 30 & \cite{Steckl96} & 45 & \cite{Chen94}%
\end{tabular}
\label{wavenumb_strong}
\end{table*}%
\normalsize

As in any other ionic crystal, the interaction of SiC with infrared light 
is clearly dominated by the fundamental vibrations of the Si- and C-sublattices 
against each other. The excitation of these phonons produces strong 
features in the wavelength range 10--13~$\mu$m, observable e.g. in 
reflectance measurements as a nearly 100\% reflectivity 
(so-called reststrahl band). The absorption
coefficient (depending on the crystal quality) can be up to several
10$^{5}$~cm$^{-1}$. Therefore, transmission measurements in this spectral 
region are only possible by using samples with effective thicknesses below 
some 100~nm.

In general, the frequencies of the involved longitudinal (LO-) and 
transverse (TO-) optical phonon modes depend on their polarization 
direction in relation to the c-axis. One reason for this behaviour - 
in simplified words - is the slight stretching of the basic
tetrahedra in c-direction causing a smaller force constant in hexagonal
double layers compared to the cubic ones (\cite{Nakashima87b}). 
One limiting case are the modes polarized parallel to the basal plane 
(perpendicular to the c-axis, E$_1$ modes), which  are very similar 
for the different polytypes. Their frequencies differ not more than 
3--4~cm$^{-1}$ (\cite{Feldman68b,Nakashima87b,Hofmann94}).
The other limiting case are the modes polarized parallel to the c-axis 
(A$_1$-modes), which have a smaller resonance frequency than the E$_1$ modes. 
For the transverse modes, the difference is nearly
proportional to the fraction of hexagonal double layers (the hexagonality
parameter h) in the polytype. This difference was experimentally found to be
$\omega(E_{1T})-\omega(A_{1T})=29.4$ cm$^{-1}\cdot$ h
(\cite{Feldman68b,Nakashima87b}) and theoretically calculated with
$\omega(E_{1T})-\omega(A_{1T})=29.2$ cm$^{-1}\cdot$ h (\cite{Hofmann94}).

Averaged values of published data of the mode frequencies of some polytypes, 
determined by Raman and infrared measurements, are given in Table \ref{wavenumb_strong}. 
For comparison, theoretical values derived from the bond charge model (\cite{Hofmann94}) 
are also given. For the 3C polytype, the maximum deviations of the single published 
frequencies from the average values in Table \ref{wavenumb_strong} are 
1.9 (TO) and 1.2 cm$^{-1}$ (LO). The biggest deviations for the E$_{1T}%
$, A$_{1T}$, E$_{1L}$, and A$_{1L}$ modes of the 6H polytype 
are 1.0, 0.9, 1.3, and 1.7 cm$^{-1}$, respectively. 

The data by \cite*{Spitzer59} for the TO- and LO mode frequencies 
of the 3C polytype, which have been often used in astronomical papers, 
seem to deviate systematically from the mean values 
(-1.9 cm$^{-1}$ and -2.0 cm$^{-1}$, respectively). 
For the E$_{1T}$ and A$_{1L}$ mode frequencies of the 6H polytype 
(\cite{Spitzer59b}), the deviations are even bigger 
(-3.1 and -4.6 cm$^{-1}$, respectively). These deviations 
probably indicate limited accuracy either of the measurements 
or of the fits in this work compared to newer ones. Therefore, 
we exclude these results from the the averaging. Including 
them would result in shifts of the mean values by an amount 
of up to 0.4 cm$^{-1}$. 

Depending on the complexity of the polytype structure, there are a number of 
weak modes in addition to the strong modes discussed so far. With increasing
unit cell length and, therefore, increasing number of atoms in the unit cell,
the number of branches in the Brillouin zone increases. Because these branches
can be folded back to the phonon dispersion curve of the basic 3C-SiC polytype
with the largest Brillouin cell (\cite{Patrick68}), these 
polytype-specific highly anisotropic modes are called zone-folded modes. They
can be observed by Raman spectroscopy (e.g. \cite{Feldman68,Nakashima97}), 
infrared reflectance (e.g. \cite{Ilin73,Pensl90}) and transmittance measurements 
with very thick samples (e.g. \cite{Dubrovskii73}). 

The peak absorption coefficient caused by these folded modes is approximately 
10$^{2}$ cm$^{-1}$. Because this is three orders of magnitude lower than the 
value for the strong bands, the weak modes are practically not observable in 
astronomical spectra.
The same is true for the combination modes of lattice vibrations 
which cause additional weak absorption effects e.g. in the spectral region 
1250--1700~cm$^{-1}$ (two-phonon combination modes, \cite{Patrick61}).

\subsection{Modelling the infrared optical properties of SiC}

SiC can be regarded as a textbook example concerning the modelling of
reflectance spectra of bulk and thin film samples
(e.g. \cite{Bohren83,Melnichuk92,Engelbrecht93}) and of transmittance spectra of thin
films (e.g. \cite{Spitzer59}) by a Lorentz-oscillator model of the
dielectric function. In many cases it is sufficient to regard only the strong
modes so that the dielectric function parallel and perpendicular
to the optical axis can be parametrized by a simple one-phonon Lorentz term: 
\begin{equation}
\varepsilon_{j}\left(  \omega\right)  =\varepsilon_{j\infty}+\ \frac
{\omega_{jP}^{2}}{\omega_{jTO}^{2}-\omega^{2}-i\ \gamma_{j}\ \omega}
\label{oszi-formel}%
\end{equation}
with
\begin{equation}
\omega_{jP}^{2}=\varepsilon_{j\infty}\ (\omega_{jLO}^{2}-\omega_{jTO}%
^{2})\qquad(j=\ \parallel,\perp).%
\end{equation}

Here the quantities $\omega_{\perp TO}$ ($\omega_{\perp LO}$) and $\omega_{\parallel TO}$
($\omega_{\parallel LO}$) are the frequencies of the E$_{1T}$ (E$_{1L}$) and
A$_{1T}$ (A$_{1L}$) modes, respectively, the plasma frequencies $\omega_{jP}$ 
represent the oscillator strengths and the $\gamma_j$ describe the phonon damping 
(line width) of the corresponding mode. The quantity $\varepsilon_{j\infty}$ 
is the high-frequency dielectric constant 
caused by the system of the valence electrons, which is measured in the visible and
NIR spectral range. Some average experimental and theoretical values of 
$\varepsilon_{j\infty}$ from the
literature are given in Table \ref{wavenumb_strong}. The scattering of the
experimental data in the literature is relatively large (up to $\pm$0.2) but
the agreement of the average values with the theoretical data by
\cite*{Chen94} is satisfactory.

The damping constant $\gamma$ (inverse proportional to the phonon life time)
is an ''ad hoc'' introduced parameter, which in a perfect crystal reflects the
anharmonicity of the potential curve. On the one hand, the phonon lifetime is
limited by coupling processes, opening energy dissipation channels. On the
other hand, phonons are scattered at crystalline defects. Therefore, the
damping constant can be taken as a measure of crystallinity. High-quality 
SiC is characterized by damping constants of 1--3~cm$^{-1}$. 
These values are reached by bulk crystals (Peiter, unpublished) as well as 
by micron-thick 
CVD-grown layers produced at relatively high temperatures (\cite{Zorba95}). 
Single-crystalline layers grown by solid-source MBE at lower temperatures 
show damping constants down to 5~cm$^{-1}$ (\cite{Pfennighaus97b}). 
Polycrystalline layers produced for instance by
sputtering or laser ablation require damping constants of several 10~cm$^{-1}$ 
in the modelling (\cite{Hobert98}). 

\subsection{Influences of real structure and impurities}

\begin{figure*}
\centering
\leavevmode
\epsfxsize=1.5
\columnwidth 
\epsfbox{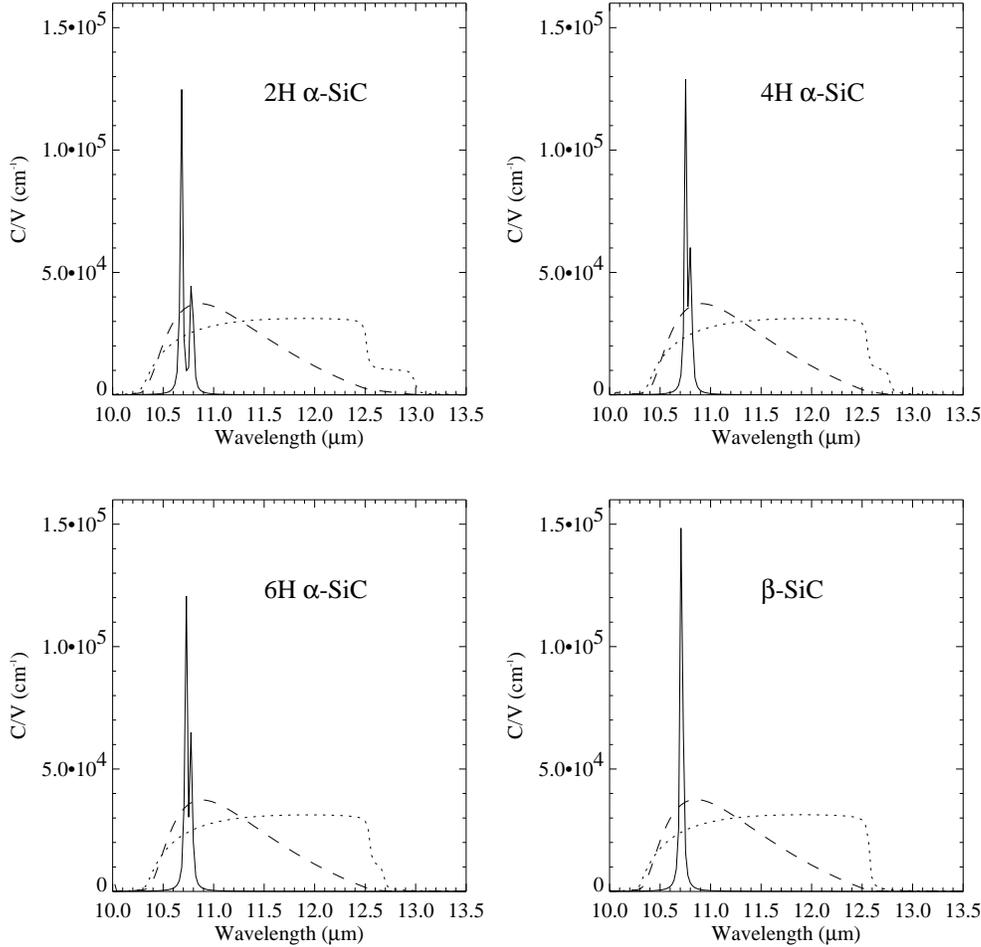}
\caption[]{Calculated absorption cross section per unit volume C/V 
for small 
SiC grains of different shapes and polytypes. Profiles for spherical grains 
are given by solid lines, the CDE according to Bohren \& Huffman (1983) by 
dotted lines, and a CDE with higher probability for less elongated grains 
(Ossenkopf et al.\,1992) by dashed lines.}
\label{calc1}
\end{figure*}

In the preceding sections more or less perfect SiC crystals have been 
considered. However, for SiC grains grown by kinetically controlled 
reactions in the outflow of carbon stars it is important 
to take the influences of structural disorder, non-stoichiometry, 
and impurities on the infrared absorption into account. 

Great experience on these effects is already available from 
thin film studies. The influence, e.g., of disorder up to complete 
amorphousness, has been investigated with films produced either by 
implantation of carbon ions into silicon wafers (e.g.  
\cite{Kimura81,Serre96}) or by amorphization of crystalline SiC by 
ion irradiation (\cite{Zorba96,Musumeci97}).
The phonon band profiles of these films in some cases still have been 
described by the Lorentz oscillator model with a very large damping 
constant of about 200~cm$^{-1}$ (\cite{Zorba96}). 
However, very often a change of the band profile from Lorentzian to Gaussian 
shape is observed (e.g. \cite{Basa90,Musumeci97}). 
The modelling of these spectra requires a Gaussian distribution of 
oscillators. 
Additionally, the centre of the oscillator frequencies shifts to smaller 
frequencies by several 10~cm$^{-1}$. The reasons for these effects 
are the variation and the (on average) increase of the bond lengths 
in the amorphous solid compared to the crystalline one. Additionally, 
the break-down of the k$\approx$0 selection rule causes contributions from 
the whole optical branches to the absorption spectrum.
Both the broadening and the shift of the absorption feature can be 
avoided or inverted by high temperature during or after the sample 
preparation, which supports recrystalization (see, e.g., \cite{Durupt82}). 

Non-stoichiometry, especially carbon deficiency, leads to the occurrence 
of separated SiC grains in some kind of matrix. The phonon band observed 
at such samples is shifted to higher frequencies 
(\cite{Yoshii91,Lundquist95}), which has been attributed to surface 
excitations of the SiC grains (\cite{Bean71}). 
In contrast to this, carbon excess leads to additional continuum 
absorption caused by the semi-metallic character of the graphitic 
excess carbon. 

Impurities and defects incorporated into the SiC crystal produce
besides a slight increase of the phonon damping (e.g. \cite{Ilin72c,Harima95}) 
the following effects: a) localized vibrational modes
(\cite{Newman90,Engelbrecht94}), b) photoionization bands of impurity
atoms (e.g. \cite{Dubrovskii71,Purtseladze72}), and c) free charge 
carrier absorption. The resulting absorption
coefficient from these impurity-induced effects is in bulk material
and films usually two or more orders of magnitude smaller than the
contributions from the strong phonons.  The free charge carrier
absorption, dominant at longer wavelength and higher temperatures,
depends on the degree of doping (\cite{Kulakowskii75}) and may become
so strong that it can be observed even in the mid-IR reflectance (carrier
density $>$\,4$\cdot$10$^{17}$cm$^{-3}$)
(\cite{Imai66,Melnichuk92}). The longitudinal
optical phonon and the free charge carrier collective oscillation
(plasmon) generally form coupled modes which occur in two branches
(L$_{+}$ and L$_{-}$) at the high and low frequency sides of the
reststrahl region (\cite{Holm86,Sasaki89}). 
The modelling of such spectra requires an additional oscillator term 
representing the response of the free charge carrier gas to electromagnetic 
excitation (so-called Drude term) in Eq.\,(\ref{oszi-formel}). 
This Drude term has a similar structure as the Lorentz term, but with 
the resonance frequency being zero and the plasma frequency being proportional 
to the square root of the charge carrier density. 

\subsection{Theoretical absorption spectra of very small spherical and 
ellipsoidal SiC grains of different polytypes}

In contrast to bulk material, the vibrational absorption bands 
of small grains are determined by surface modes which occur 
between the LO- and TO-frequencies of strong modes, the exact 
position depending on size, shape, and the medium surrounding 
the grains (\cite{Bohren83}). Within the limit of particles 
very small compared to the wavelength (quasistatic or Rayleigh 
limit), there exist simple and exact formulae for the calculation 
of absorption spectra of ellipsoidal grains from the dielectric 
function. An ensemble of such (identically shaped) grains 
embedded in random orientation in a matrix with the dielectric 
constant $\varepsilon_m$ will cause three sharp spectral features 
(surface modes) at the resonance frequencies 
$\omega_i^2=\omega_{TO}^2+L_i\omega_p^2/(\varepsilon_m+L_i(\varepsilon_\infty-\varepsilon_m))$, 
(i=1,2,3) where the geometrical factors L$_i$ (L$_i>$0, $\sum\limits_{i}$L$_i$=1) 
characterize the lengths of the principal axes of the ellipsoids 
(L$_i$=1/3 for spheres, 0, 1/2, 1/2 for needles, 0, 0, 1 for discs, 
see also \cite{Papoular98}). Unfortunately, for arbitrary, 
especially sharp-edged shapes, theory is still far from the 
ability to determine the absorption profile caused by surface 
modes of grains. A widely used possibility for providing a rough 
estimate for the absorption spectrum of a collection of real 
particles is the continuous distributions of ellipsoids (CDE) 
introduced by \cite*{Bohren83}. Besides this (rather extreme) 
uniform distribution of all possible ellipsoidal shapes 
we also use a variation of the CDE which considers a weighted 
distribution with maximum probability for spheres (m-CDE, \cite{Ossenkopf92}). 

From the dielectric function Eq.\,(\ref{oszi-formel}) with the
resonance frequencies and high-frequency dielectric constants given in
Table \ref{wavenumb_strong} we have calculated theoretical infrared 
absorption spectra for small SiC spheres as well as for the two 
CDE approximations, all in a vacuum environment. The
results of these calculations for the polytypes 3C ($\beta$-SiC), 2H,
4H, and 6H are given in Fig.\,\ref{calc1}. A damping constant
$\gamma$=2~cm$^{-1}$ which is characteristic of a good-quality crystal
(see Sect.\,2.3) was used. The orientations of the crystal axes with
respect to the polarizations of the incoming radiation have been
averaged by considering one third of the axes being oriented parallel
and two thirds normal to the polarization direction.

\begin{figure}
\centering
\leavevmode
\epsfxsize=0.8
\columnwidth 
\epsfbox{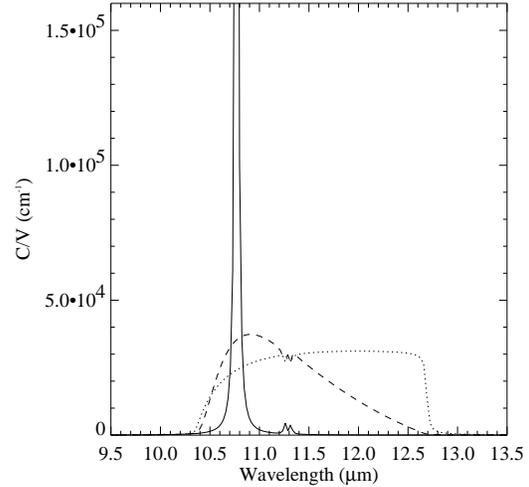}
\caption[]{Calculated absorption cross section per unit volume 
for small $\alpha$-SiC grains (6H polytype) taking into account the weak 
zone-folded modes. The lines designate different grain shape 
distributions (compare Fig.\,\ref{calc1}). For clarity only the 
polarization direction parallel to the c-axis is shown.}
\label{3oszi}
\end{figure}

\begin{figure}
\centering
\leavevmode
\epsfxsize=0.8
\columnwidth 
\epsfbox{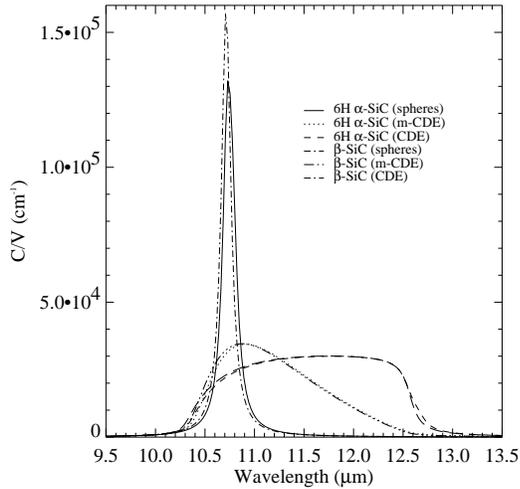}
\caption[]{Calculated absorption cross section per unit volume 
for small $\alpha$- and $\beta$-SiC grains using a higher damping constant 
of $\gamma$=10~cm$^{-1}$ than in Fig.\,\ref{calc1} (2~cm$^{-1}$).}
\label{damp}
\end{figure}

The calculated band profiles show the expected striking dependence on the 
grain shape distribution. For spherical grains very sharp resonances 
occur, whereas for the CDEs the wide distribution of shapes results in 
broad bands between the LO and TO frequencies. For the anisotropic 
modifications (the $\alpha$-SiC polytypes 2H, 4H, and 6H), the two 
principle orientations produce their 
own resonances in the sphere-spectra. 
The positions of the resonances are 10.71~$\mu$m for the 3C-, 10.73~$\mu$m
and 10.77~$\mu$m for the 6H-,\linebreak 10.75~$\mu$m and 10.80~$\mu$m for the
4H-, and 10.68~$\mu$m and 10.78~$\mu$m for the 2H-modifications, respectively. 
For comparison, the resonance positions using the data by Spitzer et al.\, 
(1959a,b) would be 10.72~$\mu$m for 3C-SiC, and 10.73~$\mu$m 
and 10.84~$\mu$m for 6H-SiC. 

\begin{table*}
\begin{center}
\caption{Properties of the investigated SiC samples. Polytype according to 
X-ray diffraction, size and shape from optical or transmission electron 
microscopy. Regular shape means observation of hexagonal/triangular 
(Dresden, Plasmachem) and spherical (Jena-Ar) grain contours. Purity 
is given according to elemental analysis (partially provided by producer); 
question marks indicate that only metal impurities have been considered. 
Major impurities are given in parentheses. Production method Acheson indicates 
high-temperature products, GPP indicates gas-phase pyrolysis (see Sect.\,3.1.)}

\begin{tabular}{llllll}
&&&&&\\
\hline\hline
product name      &polytype        &max.size&shape&purity  &prod.meth.\\
                  &                &($\mu$m)&     &(wt.\%) &\\
\hline
AJ-$\alpha$        &$\alpha$ (6H)   &2       &irr. &99.8?   &probably Acheson \\
ESK                &$\alpha$ (6H)   &2       &     &98      &Acheson \\
Lonza UF-15        &$\alpha$ (6H)   &1       &     &        &probably Acheson  \\
Duisburg           &$\alpha$ (6H)   &40      &irr. &        &probably Acheson  \\
Piesteritz (green) &$\alpha$ (6H)   &150     &irr. &98.8    &Acheson \\
Piesteritz (black) &$\alpha$ (6H,4H)&150     &irr. &96.4 (C)&Acheson \\
AJ-$\beta$         &$\beta$         &1       &irr. &96.5    &  \\
AJ whiskers        &$\beta$         &2\,x\,60&wh.  &99.5?   &  \\
Plasmachem         &$\alpha$ (6H)   &0.5     &reg. &95.2 (SiO$_2$,Si$_3$N$_4$)&plasma\\
Starck B20         &$\beta$         &0.6     &irr. &95      &  \\
Japan              &$\beta$         &0.5     &irr. &        &GPP\\
Dresden            &$\alpha$ (6H)   &0.3     &reg. &83.6 (Si$_3$N$_4$,Si)&GPP, polysilazane\\
Riga               &$\beta$         &        &     &86 (Si$_3$N$_4$,Si)&plasma\\
Berlin             &$\beta$         &0.5     &irr. &85 (C)  &plasma, SiCl$_4$\\
Jena               &$\beta$         &        &reg. &        &laser pyr., SiH$_4$,C$_2$H$_2$\\
Jena-N             &$\beta$         &        &reg. &        &laser pyr., SiH$_4$,C$_2$H$_2$,NH$_3$\\
\hline\hline
\label{mattab}
\end{tabular}
\end{center}
\end{table*}

The CDE spectra for the polytypes essentially differ in a shoulder at
the long-wavelength side of the profile, which is also caused by the
anisotropy. Since the TO-frequencies in the two principal orientations
deviate from each other, the band profile extends to different
spectral positions which results in a shoulder after the orientations
have been averaged. The difference of the E$_{1T}$ and A$_{1T}$
frequencies and consequently the broadness of the shoulder is
correlated to the hexagonality in the sense that for the 6H
modification the shoulder is only 0.14~$\mu$m broad but for the 100\%
hexagonal 2H modification its width is 0.48~$\mu$m.
In the m-CDE the shoulder is not present since the extreme shapes 
producing it have about zero probability. 

So far we have only considered the strong phonon modes. To check 
if the weak ``zone-folded'' modes could deliver a criterion useful 
for distinguishing the polytypes in small-particle spectra, we show in 
Fig.\,\ref{3oszi} as an example a calculated absorption spectrum of the 
6H polytype, which includes the weak modes. These have been represented by 
two additional oscillators with the TO- and plasma-frequencies 
$\omega_{2 TO}$=883.8~cm$^{-1}$, $\omega_{2 P}$=50~cm$^{-1}$, 
$\omega_{3 TO}$=888.6~cm$^{-1}$, $\omega_{3 P}$=45~cm$^{-1}$ 
and with $\gamma$=2~cm$^{-1}$ for both oscillators. 
The effect of the weak modes is the appearance of two very weak features 
at their TO-frequencies, interestingly as dips in the CDE spectra. 
Note that after averaging with the two times stronger contribution of the 
perpendicular orientation these features will be even weaker and hardly 
detectable in astronomical spectra. 
No influences on the resonance positions and the limiting frequencies of the 
band have been found. 

Fig.\,\ref{damp} shows calculations using a higher damping constant of
$\gamma$=10~cm$^{-1}$ which would characterize less structurally
perfect crystals but material still far from amorphousness. The calculations
have been done for the most abundant $\alpha$-SiC polytype (6H) and
for $\beta$-SiC. They show that for such less perfect crystals the
anisotropy effects in the $\alpha$-SiC spectrum are nearly smeared
out. The double feature of the spheres merged into one band with 
a peak position very close to that of the $\beta$-SiC spheres. 
The shoulder in the CDE spectrum at $\omega_{TO}$ changed into a 
smooth slope which is only a little less steep than for 
the $\beta$-SiC. 

Therefore, the result of the theoretical considerations presented in this 
section is that the small differences between the phonon frequencies 
of the SiC polytypes lead to detectable features in small-grain 
spectra only if the crystal structure of the grains is very perfect. 
Even in this case it is required that either the grains have a very definite 
shape (e.g. spherical) in order to produce sharp resonances, or that there 
is a sufficient amount of very elongated structures to produce enough 
absorption at the TO frequency so that the shoulder caused by the 
anisotropy of this frequency can be seen. The latter seems not totally 
unrealistic. However, we would like to stress that in any case the 
differences will be more marginally and that (in agreement with 
\cite{Papoular98}) the general appearance of the feature as well as 
the peak position only depend on the grain shape. 

\section{Transmission measurements on SiC particles}
\subsection{Materials and measurements}
We studied 16 different SiC powders which are partly of commercial 
origin and partly laboratory products (see Table \ref{mattab}). 
The names we assigned to these samples serve only for an easier 
distinction. For some of the commercial powders, information 
about the production route and the purity is very poor. 
Catalogue values of purity take in some cases only metal 
impurities into account and not the major pollutants: free C, Si and SiO$_2$. 

Generally, the samples fall into two groups. The first 
one is the commonly used SiC for grinding application, which is 
produced from quartz and coal in an electric furnace at very high 
temperatures (about 2500$^\circ$C, Acheson process). These powders 
are often large in grain size (Piesteritz, Duisburg) but ESK, 
Lonza and probably AJ-$\alpha$ also belong to this group. All of these 
high-temperature products consist of $\alpha$-SiC (dominated by the 
6H polytype). 

The second group is produced from gaseous or liquid precursors in some 
kind of pyrolysis which may be induced by a plasma discharge, laser 
irradiation or simply by heating. This kind of process is capable of 
producing very fine (submicron) powders and allows easy mixing with 
other components such as Si$_3$N$_4$. Therefore, a huge variety of 
such processes and products has already been studied in the literature. 
The polytype of these grains is 
usually 3C ($\beta$-SiC), but depending on the process temperature also 
$\alpha$-modifications (higher T, see Dresden, Plasmachem) or even 
amorphous components may occur. The purity is mostly not worse than that of 
the high-temperature products (if not otherwise desired as the 
Si$_3$N$_4$ component in Dresden and Riga) and the major impurities 
are the same (free C, Si, oxide). For the sake
of comparison we also have one product with a big carbon excess (Berlin). 

For astrophysically motivated studies, SiC samples from pyrolysis 
processes are certainly preferable to the high-temperature products. 
\cite*{Cherchneff97} investigated SiC particle formation in stellar 
atmospheres and showed that hydrogen must play an important role in the 
formation chemistry. In hydrogen-deficient environments as R CrB 
and Wolf-Rayet star atmospheres, SiC emission is not observed. 
The formerly proposed formation mechanism of SiC in carbon stars 
by reaction of SiC$_2$ and Si was shown by \cite*{Willacy98} 
not to be efficient enough to produce the observed quantities of SiC dust.  
If this is true, the pyrolysis processes which in nearly all cases start 
from hydrogen compounds of C and Si, will correspond much better 
to the astrophysical grain formation process. 

For all powders we determined the polytype(s) by X-ray diffraction 
as well as grain shape and size by optical or transmission electron 
microscopy (Table \ref{mattab}). For the Piesteritz products which 
have already been studied by \cite*{Friedemann81}, 
the X-ray analysis revealed a simpler composition than that supplied 
by the producer which was published by \cite*{Friedemann81}.  

Infrared spectra in the wavelength range 2--25~$\mu$m have been recorded 
with a Bruker 113v FTIR spectrometer on the SiC powders embedded in 
KBr pellets with a dilution of 0.001, which corresponds to a SiC 
column density $\sigma$\,=\,0.00015~g cm$^{-2}$. 
From the measured transmission T we derived the mass absorption 
coefficient $\kappa$\,=\,ln(1/T)/$\sigma$. 
Though this paper refers to infrared properties, it was necessary to
follow some of the spectra into the visible range. Fortunately,
KBr is transparent down to a wavelength of about 0.2~$\mu$m. Thus, 
UV/VIS spectra could be measured with the same pellets using an
Perkin Elmer Lambda 19 spectrometer. 

\subsection{Spectra of high-temperature SiC}
\begin{figure}
\centering
\leavevmode
\epsfxsize=1
\columnwidth 
\epsfbox{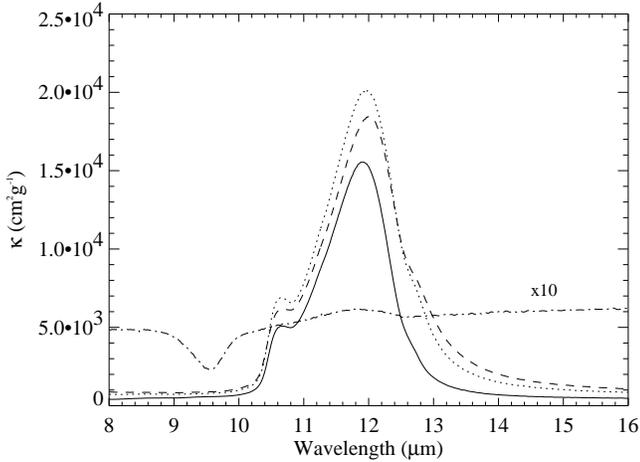}
\caption[]{Infrared band profiles of four high-temperature SiC products, 
solid line: Lonza UF-15, dashed line: ESK, dotted line: AJ alpha, 
dash-dotted line: Duisburg, enhanced by a factor of 10.}
\label{ht}
\end{figure}

Figure \ref{ht} shows the absorption spectra of four high-temperature 
products in the phonon band range. The three fine-grained powders ESK, 
Lonza UF-15 and AJ-$\alpha$ have very similar bands between about 10 
and 13.5~$\mu$m with the peak at 11.9--12.0~$\mu$m and two shoulders 
at 10.7~$\mu$m and 12.8~$\mu$m.  The maximum value of the mass 
absorption coefficient is between 15,000 and 20,000~cm$^2$g$^{-1}$. 
The profiles and absolute values are also very similar to those 
reported by \cite*{Friedemann81} and \cite*{Borghesi85}, including the 
shoulders. These have been attributed by \cite*{Friedemann81} to 
signatures of the transverse and longitudinal optical (volume) modes. 
However, in the light of Sect.\,2. we note that at least for the 
shoulder at 10.7~$\mu$m this cannot be true because the position of 
the LO mode is at shorter wavelength. The shoulders will be discussed 
further in the following sections.  Additionally we note here that 
there is only very weak continuum absorption at least at shorter 
wavelengths. To our experience, the continuum is mostly correlated to 
the degree of free carbon pollution in the sample. 

The spectrum of sample Duisburg is characterized by a more or less 
uniform level of relatively small extinction with a sharp minimum 
at 9.6~$\mu$m.  This behaviour is determined by scattering and is 
typical for large grains with diameters of 10~$\mu$m and more 
(\cite{Andersen98}). If, as indicated by studies of meteoritic SiC, 
large SiC grains occur in the interstellar medium, their different 
spectral appearance would prevent them from infrared detection. 
For the even coarser powders Piesteritz (green) and (black) with 
grain sizes of more than 100~$\mu$m, the measured spectra are 
similar to the one of the Duisburg sample but, as expected, give even 
less. Size effects will be treated further in Sect.\,3.4.

\subsection{Spectra of pyrolysis products}
\begin{figure*}
\centering
\leavevmode
\epsfxsize=1.5
\columnwidth 
\epsfbox{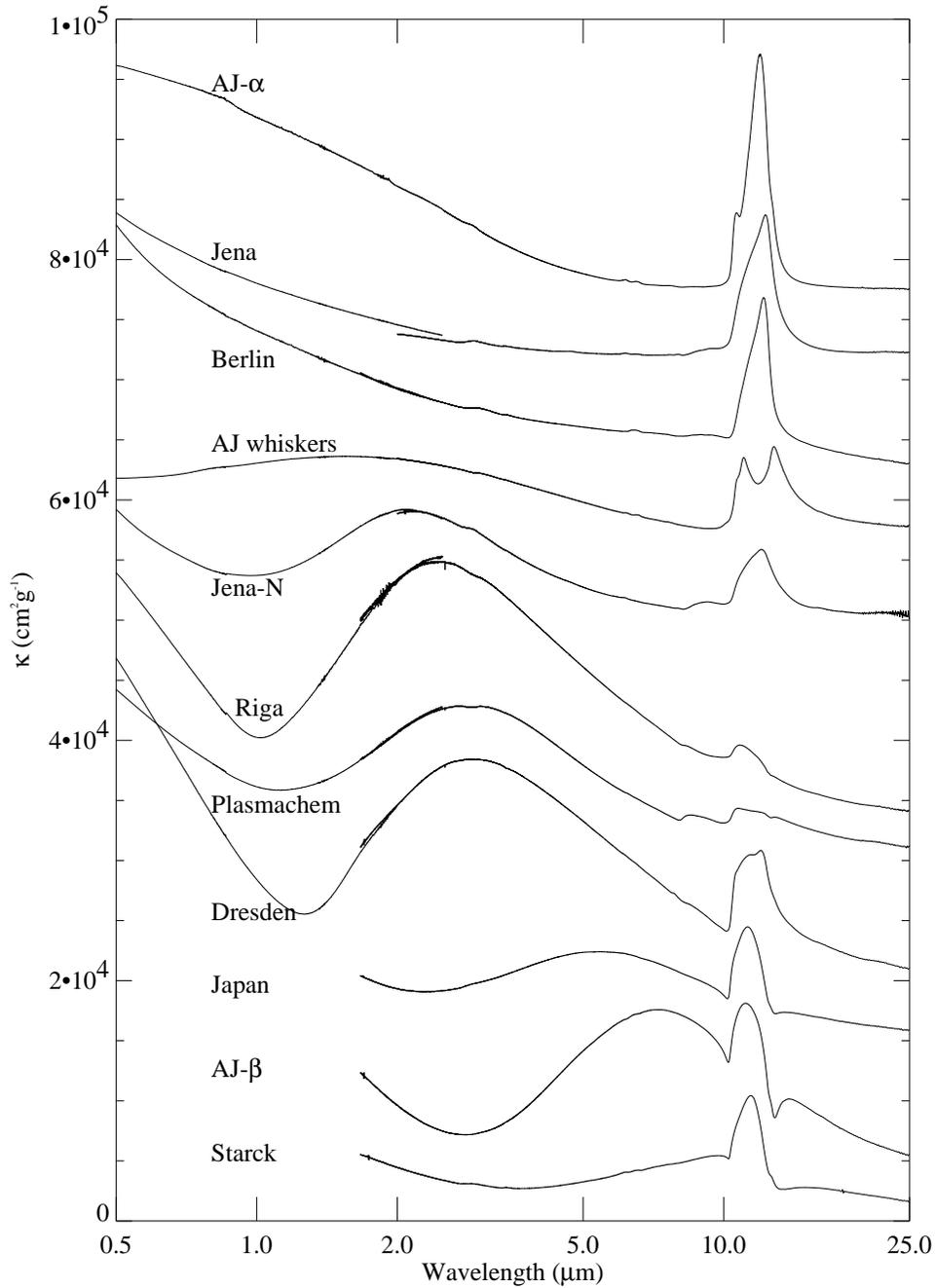}
\caption[]{VIS to IR spectra of pyrolysis SiC products (on logarithmic 
wavelength scale). 
The spectra have been arranged by vertical shifting}
\label{pyr}
\end{figure*}

Figure \ref{pyr} shows the absorption spectra of the pyrolysis products and, 
for comparison, of one high-temperature SiC. Most of the spectra have been 
measured down to a wavelength of 0.5~$\mu$m because a large part of them 
unexpectedly revealed an additional strong and extremely broad absorption 
feature, which extends in some cases from the visible range to the phonon 
band region. 
The feature partly overlaps with the phonon band and appears in first 
approximation as a kind of continuous background to the latter. 

At a closer view to the literature, one recognizes that such a broad feature 
or background is also present in the SiC spectra measured by other authors
(e.g. \cite{Borghesi85} - $\beta$-SiC, \cite{Koike98} - sample Ibiden). 
Therefore, we have to state that this kind of absorption obviously is a 
common spectral characteristics of many SiC pyrolysis products. 
In contrast to this, we did not observe the broad absorption in 
high-temperature SiC samples (see spectrum of AJ-$\alpha$). 
It certainly is not restricted to the $\beta$-polytype as is demonstrated 
by the $\alpha$-SiC samples Dresden, Plasmachem, and Ibiden (\cite{Koike98}). 

The extreme broadness of the feature excludes lattice vibrations as the 
carriers of this absorption and strongly favours collective vibrations 
of free charge carriers (plasmons). In small-particle spectra, free charge 
carrier absorption (similar to phonons) causes strong surface modes at 
frequencies smaller than the plasma frequency (\cite{Bohren83}). These 
surface modes are very well known for metal particles (\cite{Kreibig96}) 
where they occur at visible wavelengths but also for doped semiconductors 
(\cite{Rieder72,Yamamoto85}). 

The free charge carriers may have been introduced to the SiC by impurities, 
as has been already briefly discussed in Sect.\,2.4. One possibility 
is the incorporation of nitrogen which, because of its five valence electrons, 
may produce donor levels at different energy depths in the band structure 
(\cite{Chen97}). For the samples Riga, Plasmachem and Dresden the elemental 
analysis clearly detected a nitrogen content. The corresponding spectra 
are characterized by an especially high plasma frequency. Since the plasma 
frequency of a free-electron gas is proportional to the square root of the 
charge carrier density, this supports the explanation of the broad absorption 
features by nitrogen incorporation. As a test of the nitrogen hypothesis 
we prepared a nitrogen-doped SiC sample by adding a small percentage of 
NH$_3$ to the reactants. Indeed the doped sample (Jena-N) shows the 
free charge carrier absorption in contrast to the undoped one 
(Jena). These investigations will be the subject of a future paper 
(Cl\'ement et al., in prep.). 

The phonon band profiles of the pyrolysis products show a large variety 
of shapes, which are obviously not related to the polytype. The samples 
which do not possess the free charge carrier feature but only the typical 
scattering increase towards short wavelengths (Jena and Berlin), 
have phonon bands relatively similar to those of the high-temperature 
powders (compare AJ-$\alpha$). A more detailed look on the band profiles, 
however, reveals that the pyrolysis products show the band maximum 
at 12.2-12.3~$\mu$m, which is significantly longward from the 
AJ-$\alpha$ peak. Furthermore, the typical shoulders of the 
high-temperature SiC are lacking. The Jena-N sample, although it 
shows the free charge carrier absorption, has a similar phonon band 
feature. The small band at about 8--9~$\mu$m seen in this and the 
Plasmachem spectrum is due to surface oxidation. The Berlin spectrum, 
additionally, shows a background absorption of about 2000~cm$^2$g$^{-1}$ 
which is due to the free carbon excess (see Table \ref{mattab}). 
One of the most interesting phonon features is the double-peaked 
one of the AJ whiskers, which will be discussed in the following 
section. 

The phonon bands of the Riga and Plasmachem samples are very weak. 
The reason for this is totally unclear because the purity of these 
samples is better than the one of, e.g., sample Dresden. 
One possible explanation for the weakness of the phonon band 
might be the interaction with the plasmon. This interaction is rather 
obvious for the samples Japan, AJ-$\beta$, and Starck where 
the free charge carrier feature occurs at relatively large wavelengths. 
These spectra show round-shaped phonon bands peaking at very short 
wavelengths (between 11.1 and 11.4~$\mu$m). The sharp depressions 
at 10.2 and 12.8~$\mu$m limiting the spectral ranges of the phonon 
bands, and the absorption maxima at wavelengths larger than 13~$\mu$m 
point to the theoretically expected coupling of the LO phonon to the 
plasmon (\cite{Sasaki89,Digregorio93}, see also Sect.\,2.4). 

Consequently, free charge carriers in SiC grains do not only cause 
near-infrared absorption, but also determine the apparent shape 
of the phonon band if the plasmon resonance is relatively close 
to the phonon frequency. Both facts need to be considered for 
cosmic SiC grains, especially because nearly 1~at.\% of nitrogen is 
contained in the meteoritic presolar SiC (\cite{hoppe:94}). 

\subsection{Size, shape, and matrix effects}

\begin{figure}
\centering
\leavevmode
\epsfxsize=1
\columnwidth 
\epsfbox{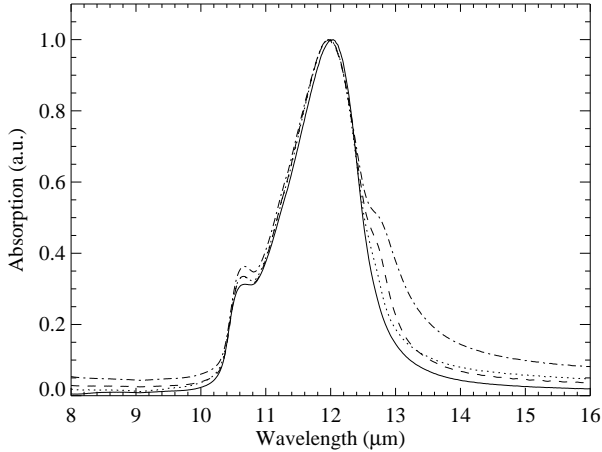}
\caption[]{Infrared band profiles of different size fractions of the product 
AJ-$\alpha$: $<$0.5~$\mu$m (solid line), 0.5--1~$\mu$m (dotted line), 
1--2~$\mu$m (dashed line) and $>$2~$\mu$m (dash-dotted line)}
\label{size}
\end{figure}

Size effects on the appearance of the SiC infrared extinction 
have been recently studied by \cite*{Andersen98} focussing on 
relatively large grains where scattering is dominating over 
absorption. We focus here on much smaller size fractions which we 
separated by sedimentation in acetone from the sample AJ-$\alpha$ 
with an original maximum grain size of 2~$\mu$m according to 
the producer. The band profiles of four size fractions are shown 
in Fig.\,\ref{size}, they may be compared with the spectrum of 
the original sample in Fig.\,\ref{ht}. The main change in the 
spectra of the different size fractions concerns the long-wavelength 
shoulder at about 12.8~$\mu$m which obviously disappears for 
the smallest size fraction. The same effect we observed 
for the other high-temperature samples. Consequently, we 
can be sure that also the shoulders found by \cite*{Friedemann81} 
and later authors around the position of the TO frequency 
indeed are due to contributions from bulk absorption 
(\cite{Bohren83}) and will vanish for very small particles. 

\begin{figure}
\centering
\leavevmode
\epsfxsize=1
\columnwidth 
\epsfbox{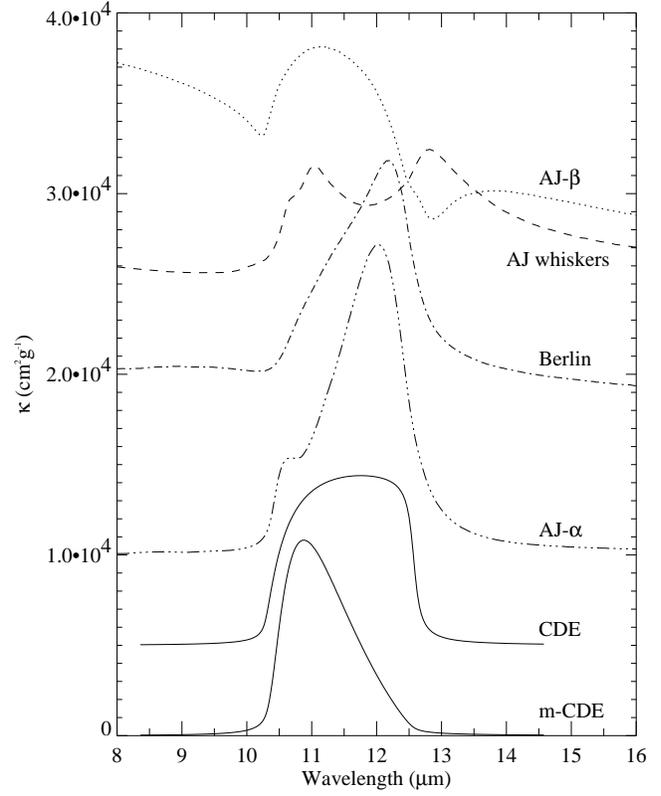}
\caption[]{Comparison of measured infrared band profiles of four 
samples (AJ-$\alpha$, Berlin, AJ whiskers, AJ-$\beta$) with 
calculated CDE profiles for $\beta$-SiC taken from Fig.\,\ref{damp}}
\label{shape}
\end{figure}

After correcting for the size effect, the phonon bands of our 
high-temperature samples nicely fit into the LO-TO interval 
10.2--12.7~$\mu$m, but still have the shoulder at about 
10.7~$\mu$m. Our calculations in Sect.\,2.5 
do not predict such a shoulder at the short-wavelength 
side of the band. In other words, it is not probable that this 
shoulder is caused by the weak anisotropy of the LO-mode frequency 
limiting the absorption band at the short-wavelength side. 
The anisotropy is much stronger for the TO frequencies 
where no shoulder is observed. 

If the optical constants of a material do not explain a feature 
of the band profile, according to Sect.~2 it must be caused by 
the grain morphology. 
Unfortunately, theory so far has no exact tool for the 
calculation of phonon band profiles for arbitrary particle 
shapes. Moreover, from the experimental side the situation 
is complicated by the fact that the SiC particles concentrate 
at the inner grain boundaries of the KBr pellet and form 
agglomerates the morphology of which is difficult to study. 
A simple approach to shape effects used e.g. by \cite*{Papoular98} 
is the approximation of real grain shapes by ellipsoids. In this 
framework any frequency in a band profile can be assigned to a 
geometrical resonance due to a special ellipsoidal shape 
characterized by the shape parameter L (see Sect.\,2.5). 
The positions of the peak and the shoulder (835 and 935~cm$^{-1}$) 
of the smallest size-fraction of Fig.\,\ref{size} would correspond 
to shape parameters of 0.08 and 0.55, respectively, which points 
to more prolate structures. This result could mean, that elongated 
agglomerates dominate in the KBr pellet. Considering the limitations 
of the ellipsoid approach, however, such an interpretation remains 
doubtful. We stress here, that for future studies of shape effects 
it is important to separate the influence of agglomeration. 

Figure \ref{shape} compares the already discussed profile of a 
high-temperature product (AJ-$\alpha$) with some typical band 
profiles of pyrolysis samples as well as with theoretical spectra 
from Sect.\,2.5. All the samples belong to the $\beta$-SiC polytype. 
The Berlin spectrum, which is also representative of the Jena 
samples, still resembles the high-temperature spectrum but peaks 
at a larger wavelength (12.2~$\mu$m) and has no distinct shoulder. 
It is the only spectrum which may be related in some way to the 
theoretical CDE spectrum, pointing to the presence of a wide 
distribution of shapes. However, it is obvious that the CDE model 
is insufficient to describe real band profiles. Therefore, 
in the case of SiC it cannot be used e.g. for the derivation of 
optical constants from IR absorption spectra of SiC particles. 
The m-CDE which initially was considered 
to be a more realistic distribution is even worse in reproducing 
the measurements, except for the AJ-$\beta$ spectrum. 
However, the band profiles of this and also the Japan and Starck 
samples (comp. Fig.\,\ref{pyr}), which all peak at very short 
wavelengths (11.1-11.4~$\mu$m), obviously are strongly influenced 
by the phonon-plasmon coupling so that a discussion in terms of 
shape effects alone is meaningless. 
It should be noted that the $\beta$-SiC spectrum published 
by \cite*{Borghesi85}, which has been used for fits of astronomical 
spectra, seems to belong to the same class. 

The most striking example of shape effects is certainly the AJ whisker 
spectrum. Similar spectra of SiC whiskers have already been 
published at different places (e.g. \cite{Pultz66,Digregorio93}). 
All these spectra are characterized by two peaks, predicted by the theory 
to occur at the TO frequency where the shape parameter is zero 
(796~cm$^{-1}$ or 12.6~$\mu$m for $\beta$-SiC) 
and at the frequency where Re($\varepsilon$)=\,-$\varepsilon_m$
(shape parameter 0.5, 929~cm$^{-1}$ or 10.8~$\mu$m in KBr) 
(\cite{Bohren83}). These positions are reproduced in our spectrum 
within 0.2~$\mu$m. The deviations probably are due to the relatively 
large size of the whiskers. 

\begin{figure}
\centering
\leavevmode
\epsfxsize=1
\columnwidth 
\epsfbox{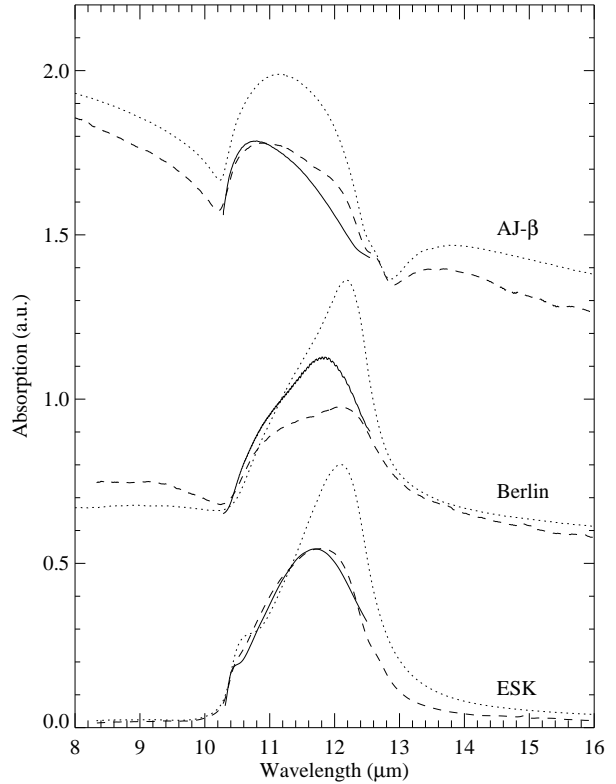}
\caption[]{Comparison of measurements on particles embedded in 
a KBr pellet (dotted lines) and infrared microscope measurements
on SiC material supported by a NaCl window (dashed lines). 
The solid curves represent the computational correction of the 
matrix effect according to the method by \cite*{Papoular98}}
\label{mat}
\end{figure}

All band profiles discussed so far have been measured with particles 
embedded in a KBr matrix which strongly influences the band shape. 
The matrix effects indeed are a big problem for comparisons of 
laboratory to astronomical spectra, since at the moment we are not 
aware of a good method to store any given powder sample in vacuum 
or in a gas for laboratory spectroscopy. Since theory is not in the 
position to calculate spectra for arbitrary grain shapes, it does 
not provide an exact method for correcting matrix effects either. 

\cite*{Papoular98} recently published the already mentioned approach, 
which is valid under the assumptions that the particles have a 
distribution of ellipsoidal shapes, and that the band can be 
described by a Lorentzian oscillator. This method works perfectly 
in test calculations with simulated spectra (if one corrects the 
spectral intensities at the calculated new resonance frequencies 
for the changed density of the latter; this step is not mentioned 
in the paper). It was also demonstrated by these authors that the 
method allows estimates of the measurement results in different 
matrices. It is clear that the predictions of this theory are 
not exact because the theory of surface phonons of real grain shapes 
(if a practicable one exists) is not the one valid for ellipsoids. 
Moreover, anisotropy still has not been included in the ellipsoid 
approach. Nevertheless, the method is useful for the interpretation 
of SiC spectra, and should definitely be used for approximate matrix 
correction instead of the formerly applied shifting of the whole 
feature. 

We have compared the predictions of Papoular's method with measurements 
which nearly avoid the matrix effect. These measurements have been 
performed by means of an infrared microscope with samples dispersed on a 
NaCl substrate (Fig.\,\ref{mat}). Thus, these grains are not fully but 
mainly surrounded by air. An important disadvantage of this method, 
which at the moment excludes it from being the solution 
to the matrix problem, is that the amount of material in the microscope 
aperture remains unknown. Therefore, the measurement is not quantitative 
but it reveals the shape of the band nearly without a matrix effect 
(compare also \cite{Andersen98}). 

For three representative samples, Fig.\,\ref{mat} displays both the 
measurements in KBr and with the microscope and, additionally, the 
matrix-corrected spectra in the interval $\omega_{LO}$--$\omega_{TO}$. 
For the calculations, the oscillator data of the 3C polytype and of 
the 6H polytype perpendicular to the c-axis (sample ESK) have been 
used since the latter dominate in a hexagonal crystal by 2:1 over 
those parallel to the c-axis. The dependence of the calculated band 
profile on these parameters (within reasonable limits) is not 
very strong. The figure shows that the microscope measurement 
confirms the blue-shift of the band center predicted by the 
calculation. This trend is expected because in a matrix of smaller 
refractive index the surface resonances have to shift to wavelengths 
where Re($\varepsilon$) of the particles is also smaller. 
The peak position and the shape of the profiles, however, are in only 
one case reproduced with satisfactory accuracy. 

Besides the already discussed inadequacy of the theory, certainly 
experimental reasons contribute to the differences. One of 
them should be the residual influence of the substrate, which 
is known from measurements on metal particles (\cite{Kreibig96}). 
However, using a silicon substrate that has a much higher 
refractive index (Si: n=3.42, NaCl: n=1.5), we found no systematic 
change of the band profiles compared to the results on NaCl substrate. 
Another reason probably is the grain clustering that 
occurs both in the KBr pellets and on the NaCl substrate, 
but may result in different morphologies. 

So at the moment, unfortunately, there is no perfect way to 
solve the matrix problem. Laboratory spectra of SiC and other 
particles with very strong vibrational or electronic bands should 
be compared to astronomical spectra only very carefully. For 
SiC we recommend the method according to \cite*{Papoular98} at 
least for a check of the possible matrix effect. Hopefully, 
also for this problem future measurements of non-agglomerated 
grains will bring more clarity. 

\section{Amorphous SiC}
\subsection{Preparation}
Amorphous SiC was obtained by the polymer pyrolysis route 
(\cite{Clement96}). 
We started from the organometallic precursor dimethylpolysilane 
(--Si(CH$_3$)$_2$)$_n$ (Strem Chemicals) which first was 
converted to polycarbosilane (PCS)
(--CH$_2$--SiHCH$_3$--CH$_2$--Si(CH$_3$)$_2$)$_n$ (\cite{Yajima76}). 
PCS reacts with oxygen and air moisture very quickly to silicon dioxide 
especially at higher temperatures. To prevent this 
reaction all preparational steps were performed in a glove box 
and the heating process under dried argon atmosphere. 

For the different analyses we have produced the amorphous SiC 
both in film and powder forms. To obtain the films, a thick sheet 
of PCS was driven on Si substrates. For the 
powders the PCS was kept in an alumina boat. The samples were 
heated slowly in a quartz tube to temperatures between 873~K and 
1323~K which were hold for one hour. 

\begin{figure}
\centering
\leavevmode
\epsfxsize=1
\columnwidth 
\epsfbox{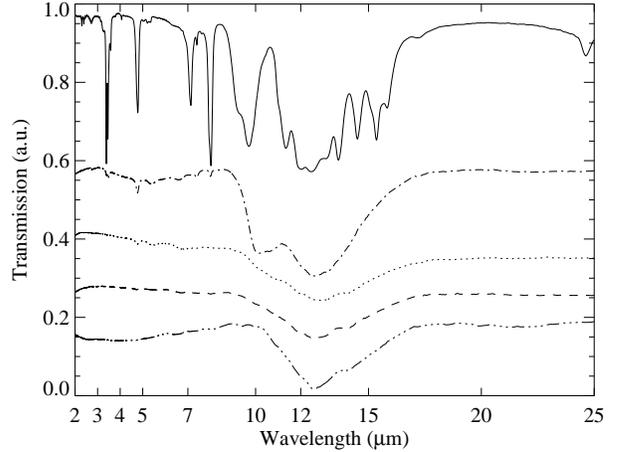}
\caption[]{Infrared band profiles of polycarbosilane (PCS) heated to 
different temperatures. From the top: 1:\,room temperature, 
2:\,873\,K, 3:\,973\,K, 4:\,1123\,K, 5:\,1273\,K.}
\label{amtemp}
\end{figure}

Figure \ref{amtemp} shows the development of the IR transmission spectra 
of the film samples with temperature. 
At temperatures lower than 1023~K the spectra still show features of the 
organic groups of the precursor PCS (see especially the feature at 
4.8~$\mu$m). Between 1073 and 1173~K, these features have vanished 
and one broad single vibrational band from 9.3 to 17.6~$\mu$m peaking 
at 12.65~$\mu$m with shoulders at about 12 and 14~$\mu$m is observed. 
X-ray diffractometry of the corresponding powder-samples revealed 
only one broad Bragg reflex in the range from 30 to 40~$^o$ 
2$\Theta$ (CuK$\alpha$ radiation). Therefore, these samples can be 
considered to be in the amorphous state. Examination by 
energy-dispersive X-ray analysis gives about the stoichiometric ratio of 
silicon to carbon atoms. 

At temperatures higher than 1223~K the infrared band becomes sharper, 
indicating the beginning crystallization of the sample (Fig.\,\ref{amtemp}). 
At 1273~K the band ranges from 10.2 to 17.6~$\mu$m with the peak 
position at the same wavelength as the amorphous samples . 
The X-ray diffractograms of these samples show narrower Bragg 
reflexes and a spike at the position of the (111) reflex of cubic SiC. 

\subsection{Optical data}
The sample annealed at 1123~K is considered to be the best amorphous 
material and is chosen for the derivation of optical constants. 
The film geometry is ideal for this purpose because it can be 
modelled more exactly than a realistic particle shape, for instance.  
Additional measurements of the produced powders could not be performed 
because the production of a sufficient volume of submicron particles 
failed.

The layer thickness of the chosen sample, which is needed for the 
derivation of the optical data from the IR transmission spectrum, 
has been determined from reflectance spectroscopy in the UV and visible 
spectral ranges. The UV/VIS spectrum shows an interference 
pattern due to multiple reflection in the SiC layer which allowes the 
simultaneous fitting of the dielectric function $\varepsilon$ 
of the layer material as well as of the layer thickness d. 
The dielectric function was modelled according to Eq.\,(\ref{oszi-formel}) 
by a constant $\varepsilon_\infty$\,=\,3.81 and by two Lorentzian oscillators 
with $\omega_{TO,1}$\,=\,48250~cm$^{-1}$, $\omega_{P,1}$\,=\,57819~cm$^{-1}$, 
$\gamma_1$\,=\,12935~cm$^{-1}$ and $\omega_{TO,2}$\,=\,36185~cm$^{-1}$, 
$\omega_{P,2}$\,=\,28492~cm$^{-1}$, $\gamma_2$\,=\,11732~cm$^{-1}$, 
respectively. 
The film thickness and the real DF in the near infrared which is 
used later as the $\varepsilon_\infty$ value for the evaluation of the 
mid-infrared transmittance spectrum were determined to be d\,=\,65 nm 
and $\varepsilon_{\infty,IR}$\,=\,5.84, respectively. The latter 
value is by about 10~\% smaller than the values for the crystalline 
modifications, which is explained by the usually smaller density of amorphous
materials.

\begin{figure}
\centering
\leavevmode
\epsfxsize=1
\columnwidth 
\epsfbox{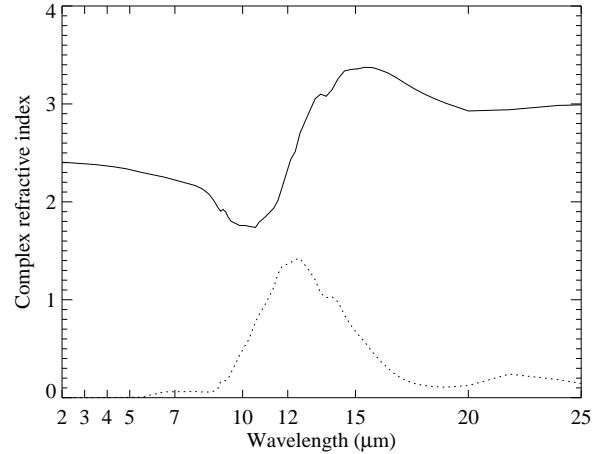}
\caption[]{The complex refractive index (real part\,=\,dotted line, imaginary part\,=\,solid 
line) calculated from the amorphous SiC layer hold at 1123 K.}
\label{amnk}
\end{figure}

\begin{figure}
\centering
\leavevmode
\epsfxsize=1
\columnwidth 
\epsfbox{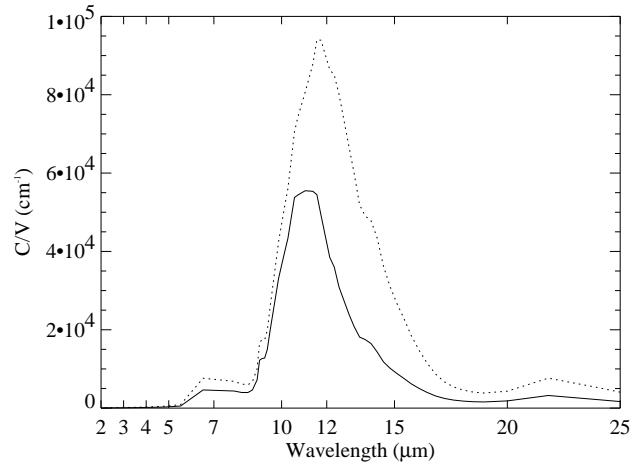}
\caption[]{The small-particle absorption spectra calculated from the optical
constants (Fig.\,\ref{amnk}): for spherical grain shape (solid line), for a 
continous distribution of ellipsoidal grain shapes (dotted).}
\label{amqabs}
\end{figure}

For the calculation of the complex refractive index from the IR 
transmission spectrum of the thin film on silicon substrate (1123K), 
a Lorentz-oscillator fit has been used, too. 
For an exact fit of the band range we applied a large number (37) of 
oscillators at different wavelength positions. 
The need for so many oscillators is a strong contrast to the properties of 
''crystalline spectra'' and reflects the presence of a wide distribution of 
vibrational modes due to the different bonding lengths in the amorphous 
material (comp. Sect.\,2.4). The resulting complex refractive index 
is shown in Fig.\,\ref{amnk}. Compared to the bands of crystalline 
materials it is, as expected, much broader. The same holds for the 
small-particle absorption spectra calculated from these optical 
constants (Fig.\,\ref{amqabs}, comp. Figs.\,\ref{ht}, \ref{pyr}, \ref{size}). 
The accompaniing ``features'' at 7 and 22~$\mu$m are artefacts caused 
by the limited accuracy of the substrate compensation. 

\section{Comparison with observations}

The goal of this paper is to give a detailed laboratory basis for an 
interpretation of astronomical SiC profiles. We do not aim at analyzing 
these profiles here, this should be the task after more high-resolution 
ISO spectra have become available. However, we will discuss the 
relevance of some of our results for astronomical infrared spectroscopy. 

If the interpretation of IRAS spectra by \cite*{Goebel95} and 
\cite*{Baron87} in terms of an a:C--H contribution is correct, 
then only the stars with relatively high continuum temperatures 
(class 4 in \cite{Goebel95}) have features of purely SiC origin. 
These features peak always around 11.3~$\mu$m. 
This is also supported by the UKIRT observations by \cite*{Speck97}, 
which are probably the pre-ISO carbon star spectra of the best quality 
so far. In these data, for continuum temperatures above 1000~K 
(e.g. of R For, IRC+50096, UU Aur, V Cyg) features dominate, which 
peak around 11.3~$\mu$m with a nearly symmetric profile and a width 
of around 1.75~$\mu$m. The profiles are not entirely symmetric, because 
there is an indication of a shoulder at about 11.8~$\mu$m. 

In contrast to this, most of our laboratory spectra show band maxima 
at relatively long wavelengths, e.g. the high-temperature products at 
11.9~$\mu$m and most pyrolysis products around 12.2~$\mu$m. Whether 
the matrix effect could have transformed band profiles peaking 
originally at 11.3~$\mu$m into the measured ones is still open, 
according to our infrared microscope measurements in Sect.\,3.4 
we rather doubt it. It is much more probable that morphological 
effects of the grains in the KBr pellets determine the measured 
laboratory profiles. However, up to now this has only been shown 
for a few samples with extreme geometries (whiskers), since at 
the moment the agglomeration state of the laboratory samples 
cannot be taken into account. 

The only laboratory samples which show the phonon features at about 
the correct position (11.1-11.4~$\mu$m) and with the correct width 
are those for which the bands are influenced by the phonon-plasmon 
coupling. To decide if the circumstellar SiC is of this type, 
there is in principle the possibility to search for an indication of the 
very broad absorption at adjacent wavelengths or for sharp minima 
limiting the phonon band region. 

Whether such an identification will be possible in the future, depends 
also on the knowledge about other features of the astronomical 
spectra, especially those caused by a-C:H and molecular species. 
Their contributions have to be known for an exact determination of 
the base line of the SiC band. Another problem of the identification 
may be self-absorption, which has already been addressed by \cite*{Speck97} 
who used it for fitting their band profiles. Self-absorption may change 
the band profile 
drastically and produce relatively broad profiles with flat tops as well 
as maxima at longer wavelengths. This might also be another possible 
explanation for the peak shift to 11.9~$\mu$m observed for the lower 
continuum temperatures, since cooler evelopes should have larger 
optical depths. Indeed, these observed features are relatively broad 
which may be difficult to explain with our data. However, they still 
are considerably narrower than the amorphous SiC feature we have 
measured (Sect.\,4). Therefore, amorphous SiC can be ruled out as an 
analog of circumstellar SiC. 

Another observational fact of importance might be the mentioned shoulder 
at about 11.8~$\mu$m. If this is not due to a:C--H absorption, it could 
be indicative of larger grain sizes of a few micron (comp. Sect.\,3.4). 
In this light, the whole observed feature shift might also reflect a 
growing grain size which would be understandable since lower continuum 
temperatures are correlated with higher mass-loss rates (\cite{skinner:88}). 
The observed weakening of the feature is also expected from a growth in 
grain size. 

\section{Conclusions}
In our discussion of the optical properties of various SiC crystal types 
in Sect.\,2 we have shown that the optical constants characterizing 
the phonon excitation by infrared radiation are too similar to cause 
significant spectral differences in small particle spectra. 
Nevertheless, in Sect.\,3 we have presented spectra of considerable 
variety which is definitely not correlated to the polytype. 
Consequently, morphology and impurities of the material are the main 
factors determining the band profile, although there may still be 
indirect effects of the polytype, e.g. via the grain shape. 
This finding opens a wide field for future studies which are 
necessary for the understanding of the cosmic silicon carbide dust. 

The points which have already been clarified or addressed in this 
paper are: 
\begin{enumerate}
\item The long-wavelength shoulder of the SiC bands measured 
by \cite*{Friedemann81} is a size effect.
\item The short-wavelength shoulder seen in these spectra is not 
positioned at $\omega_{LO}$. It may be caused by a shape effect 
but this is unclear so far. 
\item Free charge carriers strongly influence the profile of the 
phonon band via coupling to the LO phonon. 
\item The free charge carriers themselves are a source of strong 
absorption in the mid-IR. They are very common among SiC powders. 
Probably, most of the free charge carriers are produced by impurities. 
At least one of the possible dopants is nitrogen which is also 
found in the presolar meteoritic SiC grains. 
\item The crystal structure (polytype) is not a strong factor for 
the band profile, at least not in the direct way via the dielectric 
function. 
\item The infrared band profile of amorphous SiC is too wide to fit the 
11+ feature in circumstellar envelopes of carbon stars correctly.
\end{enumerate}

\begin{acknowledgements}            
The authors would like to thank Gabriele Born
and Walter Teuschel for help with the experiments and 
Dr. G. Boden, Fraunhofer-Institut Keramische Technologien und 
Sinterwerkstoffe Dresden, for providing us 
with part of the samples. We also thank the colleagues 
of the IFTO, Jena, for the permittance to reprint 
Fig.\,1. H.M. is supported by a grant of the Max Planck 
Society to Th. H. This project was partly supported 
by DFG grant Mu 1164/3-1.

\end{acknowledgements}

\bibliographystyle{aabib}
\bibliography{sic_astr,sic_lit3}
\end{document}